\documentclass[]{interact}
\usepackage{color}
\usepackage{epstopdf}
\usepackage{subfigure}
\usepackage{graphicx}
\usepackage[numbers,sort&compress]{natbib}
\bibpunct[, ]{[}{]}{,}{n}{,}{,}
\renewcommand\bibfont{\fontsize{10}{12}\selectfont}

\theoremstyle{plain}

\theoremstyle{definition}

\theoremstyle{remark}

\begin{document}


\title{Large Spontaneous {H}all Effects in Chiral Topological Magnets}

\author{
\name{S. Nakatsuji\textsuperscript{a,b}\thanks{CONTACT S. Nakatsuji. Email: satoru@issp.u-tokyo.ac.jp},  T. Higo\textsuperscript{a,b}, M. Ikhlas\textsuperscript{a}, T. Tomita\textsuperscript{a}, and Z. Tian\textsuperscript{a}}
\affil{\textsuperscript{a}Institute for Solid State Physics, The University of Tokyo, Kashiwa, Chiba 277-8581, Japan; \textsuperscript{b}CREST, Japan Science and Technology Agency (JST), 4-1-8 Honcho Kawaguchi, Saitama 332-0012, Japan.}
}

\maketitle

\begin{abstract}
As novel topological phases in correlated electron systems, we have found two examples of non-ferromagnetic states that exhibit a large anomalous Hall effect. One is the chiral spin liquid compound Pr$_2$Ir$_2$O$_7$, which exhibits a spontaneous Hall effect in a spin liquid state due to spin ice correlation. The other is the chiral antiferromagnets Mn$_3$Sn and Mn$_3$Ge that exhibit a large anomalous Hall effect at room temperature. The latter shows a sign change of the anomalous Hall effect by a small change in the magnetic field by a few 100 G, which should be useful for various applications. We will discuss that the  magnetic Weyl metal states are the origin for such a large anomalous Hall effect observed in both the spin liquid and antiferromagnet that possess almost no magnetization.
\end{abstract}


\section{Introduction}

After the discoveries of various topologically non-trivial phases in weakly correlated electron systems\cite{Hasan_TI,Fu_TI,Moore_TI,Ando,Liu14science,Neupane14,TaAs_Xu2015,TaAs_Lv2015}, it is now clear that the next place to look for novel topological phases is in correlated electron systems \cite{Kim_PRL,Machida2010,YongBaek_PRB,Leon_NP, witczak2013correlated, Leon_PRL2013}. Indeed, the hunt for such phases in correlated electron systems has been vigorously made so far. Interesting predictions that have been theoretically proposed include the Weyl semimetal phase in the iridium pyrochlore oxides\cite{Wan2011,Yang2011}, and the anomalous Hall conductors\cite{Shindou2001,Bruno2006,Martin2008,Yang2011,Ishizuka2013,Chen2014}.

Here we will discuss a chiral spin liquid \cite{Machida2010,Balicas2011,Tokiwa2013} and a chiral antiferromagnet that exhibit a large anomalous Hall effect \cite{Mn3Sn,Mn3Ge,Nayak2016} as prime candidates for novel topological phases in correlated electron systems. Conventionally, the anomalous Hall effect has been considered to be proportional to the magnetization \cite{chien,nagaosa2010}. Thus, it has never been observed in spin liquids or antiferromagnets. On the other hand, recent developments in theory to understand the anomalous Hall effect indicates that the anomalous Hall current is actually driven not by magnetization but the total of the Berry curvature of all the occupied bands\cite{nagaosa2010}. However, since the discovery of the anomalous Hall effect in 1880 \cite{Hall1880}, the effect has been seen only in ferromagnets because of the fact that the Berry curvature is usually proportional to the magnetization. On the other hand, the recent theories predict that the Berry curvature could be finite even without magnetization and in zero field, and if it is the case, there should be spin liquids and antiferromagnets that exhibit anomalous Hall effects spontaneously \cite{Shindou2001,Bruno2006,Martin2008,Yang2011,Ishizuka2013,Chen2014}. These phases in spin liquids and antiferromagnets must represent topologically nontrivial phases.

To search for such phases experimentally, we employed a strategy based on the theory for the spin chirality driven anomalous Hall effect \cite{ye,ohgushi,tatara}. The spin chirality is a solid angle subtended by neighboring three spins. According to the theory, when a conduction electron hops around these three spins, it will gain the Berry curvature proportional to the spin chirality or the solid angle. Thus, antiferromagnets with non-collinear spin texture are the best candidates that exhibit the anomalous Hall effect based on this mechanism, as opposed to ferromagnets with collinear spin order. If such a chiral antiferromagnet is found, a simple estimate shows that the Berry curvature could be as large as 100-1000 T. Obviously, the anomalous Hall effect is no longer proportional to the magnetization, and as an extreme case, the anomalous Hall effect may appear without magnetization in zero field. Such cases include i) a chiral antiferromagnet, which breaks the global time reversal symmetry, having a negligibly small magnetization \cite{Shindou2001,Bruno2006, Yang2011, Chen2014}, and ii) a chiral spin liquid, which has a spin liquid state with quantum mechanically or thermally fluctuating dipole moments, but has a hidden order that breaks time reversal symmetry such as a spin chirality order \cite{Martin2008,Ishizuka2013} and magnetic octupole \cite{Suzuki17prb}. The material class that may potentially realize these chiral phases is the geometrically frustrated magnets, and indeed we have recently found these phases in the magnetic conductors that possess pyrochlore and kagome lattices. 

\section{Large Spontaneous Hall Effect in the Spin Liquid State and a Possible Weyl Kondo Semimetal State in Pr$_2$Ir$_2$O$_7$}
The first material we discuss here is the chiral spin liquid compound, Pr$_2$Ir$_2$O$_7$ \cite{Nakatsuji2006,Machida2007,Machida2010,Balicas2011,Tokiwa2013,Kondo2015}. This material is one of the pyrochlore iridates \cite{Yanagishima,Matsuhira} and is one of the first compounds that have been recognized as a geometrically frustrated Kondo lattice \cite{Nakatsuji2006}. Namely, the material possesses two interpenetrating pyrochlore lattices, each of which provides a different electronic sector. One is the Pr based pyrochlore lattice. Each Pr, which has 4$f^2$ configuration, provides an Ising moment of $\sim3 \mu_{\rm B}$ that points along the local $<111>$ direction. They interact ferromagnetically with each other and form a spin ice state below around $T = 1.5$ K. The other is the Ir based pyrochlore lattice. The Ir$^{4+}$ state has 5$d^5$ configuration, and has mainly a $J_{\rm eff} = 1/2$ state \cite{Uematsu2015}. Notably, it has been recently found that the Ir 5$d$ bands form a Fermi node at a quadratic band touching state, or so-called Luttinger semimetal state \cite{Kondo2015}. This is the same electronic structure as the one found in HgTe \cite{HgTe_ARPES, HgTe_calc}, but with a two order of magnitude smaller band width.

\begin{figure}[t]
	\begin{center}
		\includegraphics[width=12cm]{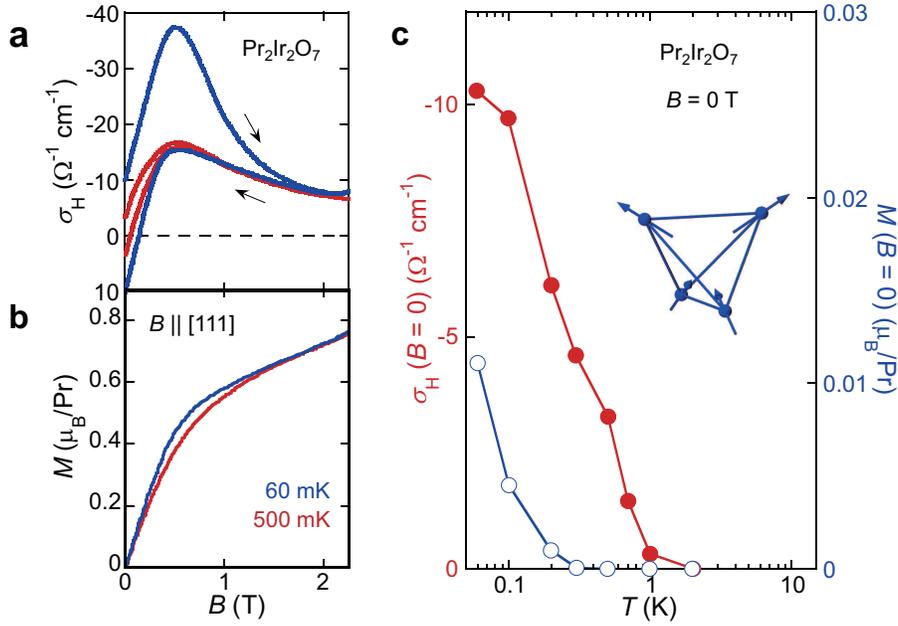}
		\caption{{\bf Spontaneous Hall effect in the spin liquid state of Pr$_2$Ir$_2$O$_7$} \cite{Machida2010} Magnetic field dependence of (a) the Hall conductivity, and (b) magnetization at 60 mK (blue) and 500 mK (red) under a field along the [111] direction. (c) Temperature dependence of the zero field spontaneous components of the Hall conductivity (open circle, left axis) and magnetization (solid circle, right axis). Each point represents a result obtained after a field cycle up to 8 T along the [111] direction.} 
	\end{center}
\end{figure}

Because of the spin ice correlation, the material does not show any magnetic long-range order down to 20 mK. The resistivity shows a minimum at $\sim 40$ K and has a rather large residual resistivity due to the low carrier density \cite{Nakatsuji2006}, consistent with the Fermi node structure\cite{Kondo2015}. While the temperature dependence of the susceptibility shows a small bifurcation between the zero-field-cooled and field-cooled curves due to a freezing of minority defect spins below around 200 mK, no magnetic order was detected in the transport, specific heat, and $\mu$SR measurements, thus indicating the ground state is a spin liquid. The material has the nearest neighbor ferromagnetic coupling scale $J \sim 0.7$ K, which has been inferred from the temperature dependence of the specific heat, susceptibility and the metamagnetic transition seen under the field along the [111] axis \cite{Machida2010}. In addition, the sister compound, Pr$_2$Zr$_2$O$_7$, also has the spin liquid state with spin ice correlation, based on the same nearest neighbor ferromagnetic coupling scale $J \sim 0.7$ K \cite{Kimura2014}. Thus, it is below $2J \sim 1.5$ K where the spin ice manifold is formed in Pr$_2$Ir$_2$O$_7$.

It is indeed in this temperature range, where a large spontaneous Hall effect, zero field anomalous Hall effect in the near absence of magnetization, has been observed \cite{Machida2010}. As can be seen in Fig. 1, the field dependence of the Hall conductivity shows a clear hysteresis near zero field and has a zero-field large intercept of 10 $\Omega^{-1}$cm$^{-1}$ at $T \sim 60$ mK (Fig. 1a). Correspondingly, the zero field anomalous Hall conductivity is found to appear below $\sim 1.5$ K, and increases on cooling and finally reaches 10 $\Omega^{-1}$cm$^{-1}$ (Fig. 1c).  Here, all the plots in Fig. 1 are reproduced from the data published in Ref. \cite{Machida2010}. Note the fact that this size of the anomalous Hall effect is so large that it can be only seen in a ferromagnet with the moment size of $\sim 1 \mu_{\rm B}$ \cite{chien,nagaosa2010}. On the other hand, the magnetization curve shows almost no hysteresis (Fig. 1b), because of the correlated paramagnetic or spin liquid state of this compound \cite{Machida2010}. As shown in Fig. 1c, the largest size of the spontaneous magnetization is $\sim 0.01 \mu_{\rm B}$, which is a few orders of magnitude smaller than that in ordinary ferromagnet. The temperature and field dependences of the Hall resistivity do not follow those of magnetization, clearly violating the empirical rule of the anomalous Hall effect, namely, $\rho_{\rm H} \propto M$. 

This large size of the anomalous Hall effect in the near absence of magnetization is striking, and calls for a new mechanism that provides a large Berry curvature in the momentum space. Taking account of the recent observation of the quadratic Fermi node \cite{Kondo2015}, one natural way of having a large Berry curvature is to have a Weyl semimetal state induced by breaking both the cubic symmetry and the global time-reversal symmetry of the quadratic band touching. To break the time reversal symmetry in a spin liquid state, the system must have a hidden order such as spin chirality order. 
If such order is directional, then it should break the cubic symmetry at the same time. Because the spin ice state has a macroscopic number of degenerate states, a local spin chirality may well form a long-range order without having any freezing in magnetic dipole moments most likely along the local high symmetry axis, namely, [111] \cite{Machida2010}. Indeed, the high anisotropy seen in the anomalous Hall effect is fully consistent with this scalar chiral order \cite{Balicas2011}. Given the fact that this system is a Kondo lattice, we may view this state as a ``Weyl Kondo semimetal'' phase, and it would be due to this Weyl semimetal state that the system exhibits the spontaneous Hall effect.

It should also be noted that this exotic phase exists nearby the putative quantum metal-insulator transition between Pr$_2$Ir$_2$O$_7$ and Nd$_2$Ir$_2$O$_7$ \cite{Matsuhira}. Nd$_2$Ir$_2$O$_7$ is the compound that shows a thermal metal-insulator transition at $T_{\rm MI} \sim 30$ K. Despite the fact that the energy gap is as large as 45 meV \cite{Ueda2012}, the insulating state can be suppressed by application of a magnetic field of around 10 T \cite{Tian2016,Ueda2015}. This type of the metal-insulator transition is very rare in this class of insulator, namely, Mott insulator \cite{Nakayama2016}. It would be interesting to find the Weyl Kondo semimetal phase nearby the field-induced quantum metal-insulator transition in Nd$_2$Ir$_2$O$_7$.

\section{Large Anomalous Hall Effect in the Chiral Antiferromagnets Mn$_3$Sn and Mn$_3$Ge}

\begin{figure}
	\begin{center}
		\includegraphics[width=\columnwidth]{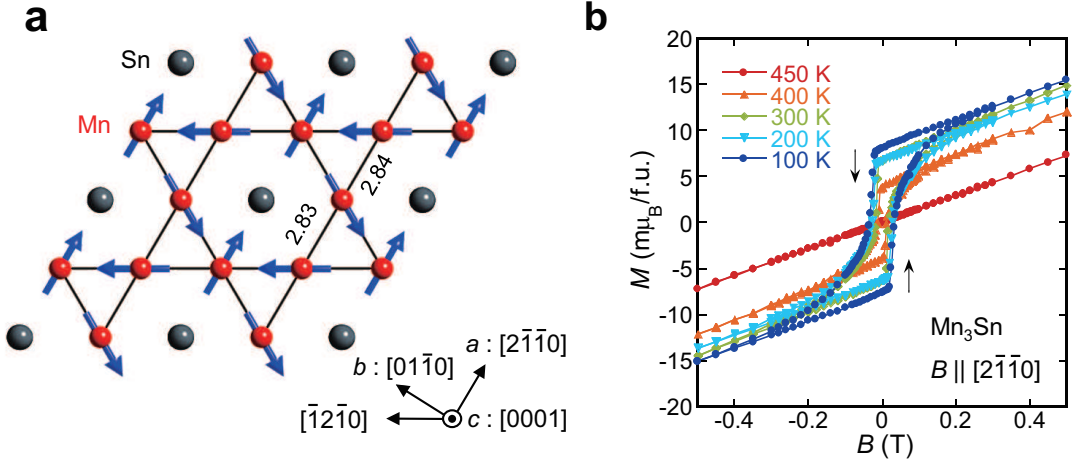}
		\caption{{\bf Crystal structure and weak ferromagnetism in Mn$_3$Sn} (a) Breathing type of Kagome structure formed by Mn atoms (small red sphere). The large gray spheres represent Sn atoms. The Mn-Mn distances are shown in \AA. The arrows indicate the Mn moments that form an inverse triangular spin structure. (b) Field dependence of the magnetization measured at various temperatures in a field along an in-plane direction.} 
	\end{center}
\end{figure}

In nature, spin liquids are much more rarely seen than antiferromagnets. Therefore, having observed the large spontaneous Hall effect in a spin liquid state\cite{Machida2010}, we thought that there should be a good chance to find an antiferromagnet that exhibits a large anomalous Hall effect. Thus, we have made an extensive search of such an antiferromagnet experimentally. Recently, we noticed that the chiral antiferromagnets Mn$_3$Sn and Mn$_3$Ge would be good candidates as they were known to exhibit weak ferromagnetism\cite{Nagamiya1982,Tomiyoshi1982,Tomiyoshi1983}, which might be due to the orbital magnetism, as we will discuss later. Indeed, we discovered these materials as the first examples of antiferromagnets that exhibit anomalous Hall effects \cite{Mn3Sn,Mn3Ge}. During our experimental study on the anomalous Hall effect in Mn$_3$Sn and Mn$_3$Ge, a theoretical work was published that discusses the anomalous Hall effects in these compounds based on first-principles calculations using spin structures that are different from those found in experiment\cite{Kubler2014}.

Mn$_3$Sn and Mn$_3$Ge are hexagonal compounds with the space group of $P6_3/mmc$ \cite{Mn3Sn,Mn3Ge}. Mn atom forms a breathing type kagome lattice (an alternating array of small and large triangles) within the $ab$-plane (Fig. 2a), which is stacked along the $c$-axis. Because of the geometrical frustration, the material shows an inverse triangular spin structure, a 120 degree spin structure with a uniform negative vector chirality \cite{Nagamiya1982,Tomiyoshi1982,Tomiyoshi1983} (Fig. 2a). In Mn$_3$Sn (Mn$_3$Ge), each Mn has a moment size of $\sim$ 3.0 (2.7) $\mu_{\rm B}$/Mn, and as a result of competition between Dzyaloshinskii--Moriya interaction and single ion anisotropy, the anisotropic energy of Mn vanishes up to four orders of spin-orbit coupling. On the other hand, the small canting of Mn moments toward a local easy axis of each Mn site leads to a weak ferromagnetic state with magnetization of $\sim$ 3 (7) m$\mu_{\rm B}$/Mn in Mn$_3$Sn (Mn$_3$Ge) \cite{Mn3Sn,Mn3Ge}. With such a small anisotropic enerygy, the coupling of the magnetic field to the weak ferromagnetism becomes essential and allows the control of the antiferromagnetic spin texture by a magnetic field.

\begin{figure}
	\begin{center}
		\includegraphics[width=\columnwidth]{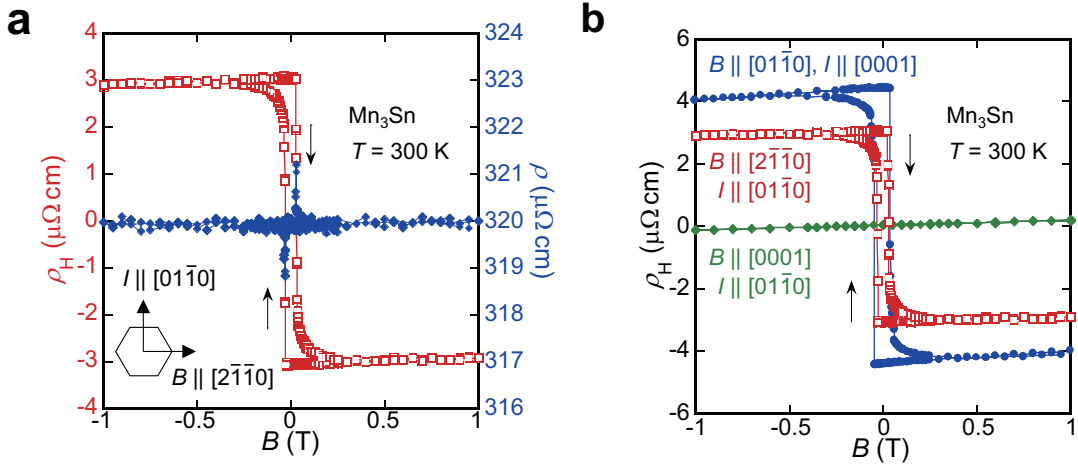}
		\caption{{\bf Large anomalous Hall effect observed at room temperature and its anisotropy in Mn$_3$Sn} (a) Magnetic field dependence of the Hall resistivity $\rho_{\rm H}(B)$ (left) and the longitudinal resistivity $\rho(B)$ in the field along an in-plane direction. The hexagon shows the hexagonal structure of Mn$_3$Sn and depicts both magnetic field ($B$) and electric current ($I$) directions. (b) Field dependence of the Hall resistivity $\rho_{\rm H}(B)$ under a field along the in-plane and out-of-plane directions.} 
	\end{center}
\end{figure}

To confirm this, we have grown single crystals of Mn$_3$Sn and Mn$_3$Ge grown by Czochralski method, and measured the magnetization curves at various temperatures \cite{Mn3Sn,Mn3Ge}. From now on, we will mainly discuss the results on Mn$_3$Sn. Figure 2b indicates the field dependence of the magnetization, which confirms that the weak ferromagnetism appears below the Neel temperature of 430 K. The observed small magnetization $\sim 3$ m$\mu_{\rm B}$/Mn is of three order of magnitude smaller than the normal spontaneous moment in a ferromagnet \cite{chien,nagaosa2010}. Thus, using the anomalous Hall coefficient known for ferromagnets, one would expect the anomalous Hall resistivity of $\sim$ 0.01 $\mu\Omega$cm for this antiferromagnetic state. However, as we will see, what we discovered here is $\sim$ 3 $\mu\Omega$cm, which is 300 times larger than the naively estimated value based on the magnetization\cite{Mn3Sn,Mn3Ge}. This is extremely large, and cannot be explained by the size of the magnetization and thus indicates unusual mechanism at work.

In fact, as shown in Fig. 3a, the Hall resistivity shows a sharp jump of $\sim$ 6 $\mu\Omega$cm with a small coercivity of $\sim$ 100 G. Using the estimate of the normal Hall coefficient of $R_{\rm 0} < 0.015 \ \mu \Omega$cm/T, the field necessary to induce this size of the Hall resistivity is more than 200 T. This indicates the fictitious field of more than $\sim$ 200 T must exist in the momentum space of the system. On the other hand, the magnetoresistance is almost constant up to 1 T, which make it very easy to detect and estimate the Hall resistivity component (Fig. 3a). Furthermore, the spontaneous Hall resistivity shows anisotropic character that is fully consistent with the anisotropy of the magnetization (Fig. 3b). Namely, it only appears in the field along the $ab$-plane, but no spontaneous component was seen in the field along the $c$-axis. 

\begin{figure}
	\begin{center}
		\includegraphics[width=7cm]{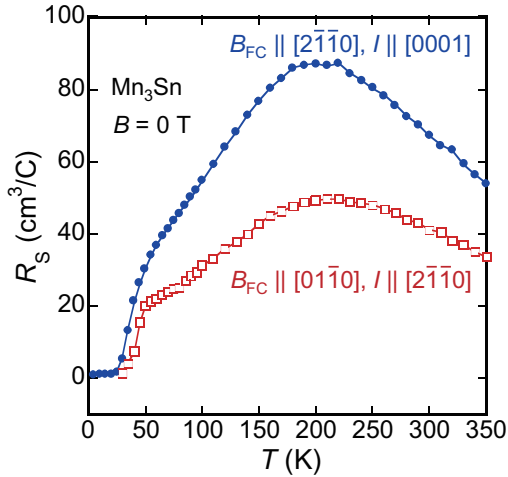}
		\caption{{\bf Temperature dependence of the anomalous Hall coefficient in Mn$_3$Sn} Temperature dependence of the anomalous Hall coefficient $R_{\rm s} = \rho_{\rm H}(B=0)/M(B=0)$ obtained after a field cooling procedures in the field of $B_{\rm FC} = 7$ T. The directions for the field used for the field cooling and electric current are shown in the figure. Here, $\rho_{\rm H}(B=0)$, and $M(B=0)$ are the zero field components of the Hall resistivity and the magnetization.} 
	\end{center}
\end{figure}

Empirically, the anomalous Hall resistivity is known to be proportional to magnetization $M$, and the coefficient $R_{\rm s} = \rho_{\rm H}/M$ provides a measure of the sensitivity of the anomalous Hall effect to magnetization\cite{hurd,chien}. This usually takes the value of the order of $10^{-4}-10^{-2}$ cm$^3$/C for ferromagnets such as Fe, Co, Ni, and Permalloy\cite{chien,nagaosa2010}. To estimate $R_{\rm s}$, we measured the spontaneous magnetization $M(B=0)$ and Hall conductivity $\sigma_{\rm H}(B=0)$ using a field-cooling procedure. Figure 4 shows the temperature dependence of $R_{\rm s} = |\rho_{\rm H}(B=0)/M(B=0)|$ for Mn$_3$Sn. In comparison with the values known for metallic ferromagnets, $R_{\rm s}$ for $B \parallel [2\bar{1}\bar{1}0]$ and $[01\bar{1}0]$ are extremely large by a factor of $>10^4$ and indicates an unusual mechanism of the anomalous Hall effect in Mn$_3$Sn.

The observed large anomalous Hall effect in antiferromagnets is surprising and requires a special explanation for the mechanism. For the possible anomalous Hall effect in another non-collinear antiferromagnet Mn$_3$Ir, whose structure can be viewed as two dimensional kagome lattice stacked along the [111] direction, Chen {\it et al.} proposes that the anomalous Hall effect may be induced by the breaking of the $TM$ symmetry by the spin-orbit coupling, i.e. the combination of the time-reversal symmetry $T$ and the mirror symmetry $M$ of the kagome lattice, which is otherwise a good symmetry of a single layer kagome lattice \cite{Chen2014}. 
A similar symmetry argument applies to Mn$_3$Sn and the fact that magnetic ordering lowers the symmetry from hexagonal to orthorhombic one allows the system to have an anomalous Hall effect in the $ab$ plane. Indeed, the magnetic moments on two neighboring triangles of Mn forming an octahedron can be viewed as a magnetic octupole and thus the magnetic structure of Mn$_3$Sn has a uniform ferroic order of this magnetic octupole, breaking the time-reversal symmetry macroscopically \cite{Suzuki17prb}. In addition, the numerical calculation by K\"ubler {\it et al.} has revealed a large anomalous Hall effect in Mn$_3$Sn \cite{Kubler2014}, while they used the magnetic structure that is different from the one observed in experiment\cite{Nagamiya1982,Tomiyoshi1982,Tomiyoshi1983}. The anomalous Hall conductivity can be computed by the Brillouin zone integration of the Berry curvature of all occupied bands and they found a significant enhancement of the Berry curvature particularly around the band crossing points called Weyl points \cite{Wan2011,Burkov2011} near the Fermi energy. Using the spin structure that has been reported in the experimental literature, a recent first principles calculation has confirmed that both Mn$_3$Sn and Mn$_3$Ge should be Weyl metals and have the large Berry curvature due to the Weyl points near the Fermi energy to enhance the anomalous Hall effect \cite{Yang2017njp}. Up to now, no confirmation has been made for the magnetic Weyl semimetal, and it is a significant subject to experimentally confirm the existence of the Weyl fermions for example by ARPES measurements and magnetotransport studies to observe the chiral anomaly \cite{Nielsen83plb,Son2013,Xiong2015,Zhang2016,Hirschberger16naturemat}. In addition, this system should serve as a model system to reveal the elusive effects of Berry curvature, as the large anomalous Hall effect observed at room temperature should allow us to find novel phenomena induced by the Berry curvature using various tools that are available at room temperature.

In addition, the spontaneous zero field part of the anomalous Hall resistivity has qualitatively different response to the magnetization in comparison with the magnetic field induced part \cite{Mn3Sn,Mn3Ge}. This leads us to speculate that the weak FM component comes not only from the canting of the sublattice moments as in the high field regime, but from the orbital magnetization that originate from the Berry curvature that induces the large anomalous Hall effect \cite{Fukuyama,Niu2010}. The orbital magnetization due to the Berry curvature has been discussed for various theoretical models that predict a large anomalous Hall effect in antiferromagetic states \cite{Shindou2001,Martin2008}. In comparison with ferromagnets, antiferromagnets have much smaller spin magnetization, and therefore should provide better place to confirm the elusive weak magnetization produced by itinerant circulating current at the edge of bulk samples. It is thus interesting to examine the origin of the weak FM magnetization, in particular, the possibility of orbital magnetization.

In recent years, there is a surge in the interest in antiferromagnets in the field of spintronics \cite{MacDonald2006, Jungwirth2010,MacDonald2011,Park2011,Marti2014,Gomonay2014, Jungwirth2016}. This is because antiferromagnets usually have almost no stray field that perturbs neighboring cells and therefore, denser integration is possible by using antiferromagnets, in sharp contrast with ferromagnets that have been used as a main active component\cite{Fert2007}. Moreover, antiferromagnets have weak sensitivity to the magnetic field and thus provide good data retention. Plus, much faster dynamics in antiferromagnets than in ferromagnets is useful for ultrafast data processing. On top, Mn$_3$Sn is a stable and easily fabricable compound made of low cost elements. Thus, our finding that the large fictitious field of more than $\sim 200$ T can be controlled by a few 100 G places Mn$_3$Sn as a potential material that is useful for future spintronics application, including memory device and optical switching.

\section{Summary}
Here we discussed the large spontaneous Hall effects observed in both a spin liquid and antiferromagnetic states. In sharp contrast with the conventional anomalous Hall effect in ferromagnets, the effects with almost no magnetization are obviously unusual, and clearly indicate topologically non-trivial electronic structures as their origin. Our recent observations in addition to theoretical studies indeed point to the Weyl Kondo semimetal state in Pr$_2$Ir$_2$O$_7$ and the Weyl metal state in Mn$_3$Sn and Mn$_3$Ge. Their topologically robust properties against small perturbation such as impurity doping or disorder will be useful in their future technological application. Obviously, our findings represent the tip of the iceberg of the fascinating sets of discoveries that will be coming in the near future in the study of this new class of topological magnets.

\section*{Acknowledgements}
We thank Naoki Kiyohara, Yosuke Matsumoto, Akito Sakai, Collin Broholm, Luis Balicas, Takashi Kondo, Ryotaro Arita, Michito Suzuki, Leon Balents and Yong Baek Kim for useful discussions.
This work is partially supported by CREST (JPMJCR15Q5), Japan Science and Technology Agency, by Grants-in-Aid for Scientific Research (16H02209), by Grants-in-Aids
for Scientific Research on Innovative Areas (15H05882, 15H05883) and Program for Advancing Strategic International Networks to Accelerate the Circulation of Talented Researchers (No. R2604) from the Japanese Society for the Promotion of Science. The use of the facilities of the Materials Design and Characterization Laboratory at the Institute for Solid State Physics, The University of Tokyo, is gratefully acknowledged.

\bibliographystyle{tfq}

\begin{thebibliography}{10}
	\newcommand{\printfirst}[2]{#1}
	\newcommand{\switchargs}[2]{#2#1}
	\providecommand{\url}[1]{\normalfont{#1}}
	\providecommand{\urlprefix}{Available at }
	
	\bibitem{Hasan_TI}
	M.Z. Hasan and C.L. Kane, \emph{Colloquium: {T}opological insulators}, Rev.
	Mod. Phys. 82 (2010), p. 3045.
	
	\bibitem{Fu_TI}
	L. Fu, C.L. Kane, and E.J. Mele, \emph{Topological {I}nsulators in {T}hree
		{D}imensions}, Phys. Rev. Lett. 98 (2007), p. 106803.
	
	\bibitem{Moore_TI}
	J.E. Moore and L. Balents, \emph{Topological invariants of
		time-reversal-invariant band structures}, Phys. Rev. B 75 (2007), p. 121306.
	
	\bibitem{Ando}
	Y. Ando, \emph{Topological {I}nsulator {M}aterials}, J. Phys. Soc. Jpn. 82
	(2013), p. 102001.
	
	\bibitem{Liu14science}
	Z.K. Liu, B. Zhou, Y. Zhang, Z.J. Wang, H.M. Weng, D. Prabhakaran, S.K. Mo,
	Z.X. Shen, Z. Fang, X. Dai, Z. Hussain, and C.Y. L., \emph{Discovery of a
		three-dimensional topological {Dirac} semimetal, {Na}$_3${Bi}}, Science 343
	(2014), pp. 864--867.
	
	\bibitem{Neupane14}
	M. Neupane, S.Y. Xu, R. Sankar, N. Alidoust, G. Bian, C. Liu, I. Belopolski,
	T.R. Chang, H.T. Jeng, and H. Lin, \emph{Observation of a three-dimensional
		topological {D}irac semimetal phase in high-mobility {C}d{$_3$}{A}s{$_2$}},
	Nature commun. 5 (2014).
	
	\bibitem{TaAs_Xu2015}
	S.Y. Xu, I. Belopolski, N. Alidoust, M. Neupane, G. Bian, C. Zhang, R. Sankar,
	G. Chang, Z. Yuan, C.C. Lee, S.M. Huang, H. Zheng, J. Ma, D.S. Sanchez, B.
	Wang, A. Bansil, F. Chou, P.P. Shibayev, H. Lin, S. Jia, and M.Z. Hasan,
	\emph{Discovery of a {W}eyl fermion semimetal and topological {F}ermi arcs},
	Science .
	
	\bibitem{TaAs_Lv2015}
	B.Q. Lv, H.M. Weng, B.B. Fu, X.P. Wang, H. Miao, J. Ma, P. Richard, X.C. Huang,
	L.X. Zhao, G.F. Chen, Z. Fang, X. Dai, T. Qian, and H. Ding,
	\emph{Experimental {D}iscovery of {W}eyl {S}emimetal {T}a{A}s}, Phys. Rev. X
	5 (2015), p. 031013.
	
	\bibitem{Kim_PRL}
	B.J. Kim, H. Jin, S.J. Moon, J.Y. Kim, B.G. Park, C.S. Leem, J. Yu, T.W. Noh,
	C. Kim, S.J. Oh, J.H. Park, V. Durairaj, G. Cao, and E. Rotenberg,
	\emph{Novel {J}$_{eff}$= 1/2 {M}ott {S}tate {I}nduced by {R}elativistic
		{S}pin-{O}rbit {C}oupling in {S}r$_2${I}r{O}$_4$}, Phys. Rev. Lett. 101
	(2008), p. 076402.
	
	\bibitem{Machida2010}
	Y. Machida, S. Nakatsuji, S. Onoda, T. Tayama, and T. Sakakibara,
	\emph{Time-reversal symmetry breaking and spontaneous {H}all effect without
		magnetic dipole order}, Nature 463 (2010), pp. 210--213.
	
	\bibitem{YongBaek_PRB}
	W. Witczak-Krempa and Y.B. Kim, \emph{Topological and magnetic phases of
		interacting electrons in the pyrochlore iridates}, Phys. Rev. B 85 (2012), p.
	045124.
	
	\bibitem{Leon_NP}
	D. Pesin and L. Balents, \emph{Mott physics and band topology in materials with
		strong spin-orbit interaction}, Nature Phys. 6 (2010), pp. 376--381.
	
	\bibitem{witczak2013correlated}
	W. Witczak-Krempa, G. Chen, Y.B. Kim, and L. Balents, \emph{Correlated
		{Q}uantum {P}henomena in the {S}trong {S}pin-{O}rbit {R}egime}, Annu. Rev.
	Condens. Matter Phys. 5 (2014), pp. 57--82.
	
	\bibitem{Leon_PRL2013}
	E.G. Moon, C. Xu, Y.B. Kim, and L. Balents, \emph{{N}on-{F}ermi-{L}iquid and
		{T}opological {S}tates with {S}trong {S}pin-{O}rbit {C}oupling}, Phys. Rev.
	Lett. 111 (2013), p. 206401.
	
	\bibitem{Wan2011}
	X. Wan, A.M. Turner, A. Vishwanath, and S.Y. Savrasov, \emph{Topological
		semimetal and {Fermi}-arc surface states in the electronic structure of
		pyrochlore iridates}, Phys. Rev. B 83 (2011), p. 205101.
	
	\bibitem{Yang2011}
	K.Y. Yang, Y.M. Lu, and Y. Ran, \emph{Quantum {H}all effects in a {W}eyl
		semimetal: {P}ossible application in pyrochlore iridates}, Phys. Rev. B 84
	(2011), p. 075129.
	
	\bibitem{Shindou2001}
	R. Shindou and N. Nagaosa, \emph{Orbital {F}erromagnetism and {A}nomalous
		{H}all {E}ffect in {A}ntiferromagnets on the {D}istorted fcc {L}attice},
	Phys. Rev. Lett. 87 (2001), p. 116801.
	
	\bibitem{Bruno2006}
	G. Metalidis and P. Bruno, \emph{Topological {Hall} effect studied in simple
		models}, Phys. Rev. B 74 (2006), p. 045327.
	
	\bibitem{Martin2008}
	I. Martin and C.D. Batista, \emph{{I}tinerant {E}lectron-{D}riven {C}hiral
		{M}agnetic {O}rdering and {S}pontaneous {Q}uantum {Hall} {E}ffect in
		{T}riangular {L}attice {M}odels}, Phys. Rev. Lett. 101 (2008), p. 156402.
	
	\bibitem{Ishizuka2013}
	H. Ishizuka and Y. Motome, \emph{Quantum anomalous {H}all effect in kagome
		ice}, Phys. Rev. B 87 (2013), p. 081105.
	
	\bibitem{Chen2014}
	H. Chen, Q. Niu, and A.H. MacDonald, \emph{Anomalous {H}all {E}ffect {A}rising
		from {N}oncollinear {A}ntiferromagnetism}, Phys. Rev. Lett. 112 (2014), p.
	017205.
	
	\bibitem{Balicas2011}
	L. Balicas, S. Nakatsuji, Y. Machida, and S. Onoda, \emph{Anisotropic
		{H}ysteretic {H}all {E}ffect and {M}agnetic {C}ontrol of {C}hiral {D}omains
		in the {C}hiral {S}pin {S}tates of {P}r$_2${I}r$_2${O}$_7$}, Phys. Rev. Lett.
	106 (2011), p. 217204.
	
	\bibitem{Tokiwa2013}
	Y. Tokiwa, J.J. Ishikawa, S. Nakatsuji, and P. Gegenwart, \emph{Quantum
		criticality in a metallic spin liquid}, Nature Mat. 13 (2014), pp. 356--359.
	
	\bibitem{Mn3Sn}
	S. Nakatsuji, N. Kiyohara, and T. Higo, \emph{{L}arge {A}nomalous {H}all
		{E}ffect in a {N}on-collinear {A}ntiferromagnet at {R}oom {T}emperature},
	Nature 527 (2015), pp. 212--215.
	
	\bibitem{Mn3Ge}
	N. Kiyohara, T. Tomita, and S. Nakatsuji, \emph{Giant {A}nomalous {H}all
		{E}ffect in the {C}hiral {A}ntiferromagnet {Mn}$_{3}${Ge}}, Phys. Rev.
	Applied 5 (2016), p. 064009.
	
	\bibitem{Nayak2016}
	A.K. Nayak, J.E. Fischer, Y. Sun, B. Yan, J. Karel, A.C. Komarek, C. Shekhar,
	N. Kumar, W. Schnelle, J. K{\"u}bler, C. Felser, and S.S.P. Parkin,
	\emph{Large anomalous {H}all effect driven by a nonvanishing {B}erry
		curvature in the noncolinear antiferromagnet {Mn$_3$Ge}}, Sci. Adv. 2 (2016),
	p. e1501870.
	
	\bibitem{chien}
	C.L. Chien and C.R. Westgate, \emph{The {H}all Effect and its Applications},
	Plenum, New York, 1980.
	
	\bibitem{nagaosa2010}
	N. Nagaosa, J. Sinova, S. Onoda, A.H. MacDonald, and N.P. Ong, \emph{Anomalous
		{H}all effect}, Rev. Mod. Phys. 82 (2010), pp. 1539--1592.
	
	\bibitem{Hall1880}
	E.H. Hall, \emph{On the ``{R}otational {C}oefficient'' in {N}ickel and
		{C}obalt}, Proc. Phys. Soc. Lond. 4 (1880), pp. 325--342.
	
	\bibitem{ye}
	J. Ye, Y.B. Kim, A.J. Millis, B.I. Shraiman, P. Majumdar, and Z. Tesanovic,
	\emph{Berry phase theory of the anomalous {Hall} effect: application to
		colossal magnetoresistance manganites}, Phys.\ Rev.\ Lett. 83 (1999), p.
	3737.
	
	\bibitem{ohgushi}
	K. Ohgushi, S. Murakami, and N. Nagaosa, \emph{Spin anisotropy and quantum
		{Hall} effect in the kagome lattice: {Chiral} spin statebased on a
		ferromagnet}, Phys.\ Rev.\ B 62 (2000), p. R6065.
	
	\bibitem{tatara}
	G. Tatara and H. Kawamura, \emph{Chirality-{D}riven {A}nomalous {Hall} {E}ffect
		in {W}eak {C}oupling {R}egime}, J.\ Phys.\ Soc.\ Jpn. 71 (2002), p. 2613.
	
	\bibitem{Suzuki17prb}
	M.T. Suzuki, T. Koretsune, M. Ochi, and R. Arita, \emph{Cluster multipole
		theory for anomalous {Hall} effect in antiferromagnets}, Phys. Rev. B 95
	(2017), p. 094406.
	
	\bibitem{Nakatsuji2006}
	S. Nakatsuji, Y. Machida, Y. Maeno, T. Tayama, T. Sakakibara, J. van  Duijn, L.
	Balicas, J.N. Millican, R.T. Macaluso, and J.Y. Chan, \emph{Metallic
		{S}pin-{L}iquid {B}ehavior of the {G}eometrically {F}rustrated {K}ondo
		{L}attice {P}r$_2${I}r$_2${O}$_7$}, Phys. Rev. Lett. 96 (2006), p. 087204.
	
	\bibitem{Machida2007}
	Y. Machida, S. Nakatsuji, Y. Maeno, T. Tayama, T. Sakakibara, and S. Onoda,
	\emph{Unconventional {A}nomalous {H}all {E}ffect {E}nhanced by a
		{N}oncoplanar {S}pin {T}exture in the {F}rustrated {K}ondo {L}attice
		{P}r$_2${I}r$_2${O}$_7$}, Phys. Rev. Lett. 98 (2007), p. 057203.
	
	\bibitem{Kondo2015}
	T. Kondo, M. Nakayama, R. Chen, J. Ishikawa, E.G. Moon, T. Yamamoto, Y. Ota, W.
	Malaeb, H. Kanai, Y. Nakashima, Y. Ishida, R. Yoshida, H. Yamamoto, M.
	Matsunami, S. Kimura, N. Inami, K. Ono, H. Kumigashira, S. Nakatsuji, L.
	Balents, and S. Shin, \emph{Quadratic {F}ermi node in a 3{D} strongly
		correlated semimetal}, Nature Commun. 6 (2015), p. 10042.
	
	\bibitem{Yanagishima}
	D. Yanagishima and Y. Maeno, \emph{Metal-{N}onmetal {C}hangeover in
		{P}yrochlore {I}ridates}, J. Phys. Soc. Jpn. 70 (2001), p. 2880.
	
	\bibitem{Matsuhira}
	K. Matsuhira, M. Wakeshima, R. Nakanishi, T. Yamada, A. Nakamura, W. Kawano, S.
	Takagi, and Y. Hinatsu, \emph{Metal-{I}nsulator {T}ransition in {P}yrochlore
		{I}ridates {L}n$_2${I}r$_2${O}$_7$ ({L}n= {N}d, {S}m, and {E}u)}, J. Phys.
	Soc. Jpn. 76 (2007), p. 3706.
	
	\bibitem{Uematsu2015}
	D. Uematsu, H. Sagayama, T.h. Arima, J.J. Ishikawa, S. Nakatsuji, H. Takagi, M.
	Yoshida, J. Mizuki, and K. Ishii, \emph{Large trigonal-field effect on
		spin-orbit coupled states in a pyrochlore iridate}, Phys. Rev. B 92 (2015),
	p. 094405.
	
	\bibitem{HgTe_ARPES}
	C. Br{\"u}ne, C.X. Liu, E.G. Novik, E.M. Hankiewicz, H. Buhmann, Y.L. Chen,
	X.L. Qi, Z.X. Shen, S.C. Zhang, and L.W. Molenkamp, \emph{Quantum {H}all
		effect from the topological surface states of strained bulk {H}g{T}e}, Phys.
	Rev. Lett. 106 (2011), p. 126803.
	
	\bibitem{HgTe_calc}
	S. Zaheer, S.M. Young, D. Cellucci, J.C.Y. Teo, C.L. Kane, E.J. Mele, and A.M.
	Rappe, \emph{Spin texture on the {F}ermi surface of tensile-strained
		{H}g{T}e}, Phys. Rev. B 87 (2013), p. 045202.
	
	\bibitem{Kimura2014}
	K. Kimura, S. Nakatsuji, J.J. Wen, C. Broholm, M.B. Stone, E. Nishibori, and H.
	Sawa, \emph{Quadratic {F}ermi node in a 3{D} strongly correlated semimetal},
	Nature Commun. 4 (2013), p. 2914.
	
	\bibitem{Ueda2012}
	K. Ueda, J. Fujioka, Y. Takahashi, T. Suzuki, S. Ishiwata, Y. Taguchi, and Y.
	Tokura, \emph{Variation of charge dynamics in the course of metal-insulator
		transition for pyrochlore-type {N}d$_{2}${I}r$_2${O}$_{7}$}, Phys. Rev. Lett.
	109 (2012), p. 136402.
	
	\bibitem{Tian2016}
	Z. Tian, Y. Kohama, T. Tomita, H. Ishizuka, T.H. Hsieh, J.J. Ishikawa, K.
	Kindo, L. Balents, and S. Nakatsuji, \emph{Field-induced quantum
		metal-insulator transition in the pyrochlore iridate
		{Nd$_{2}$Ir$_2$O$_{7}$}}, Nature Phys. 12 (2016), pp. 134--138.
	
	\bibitem{Ueda2015}
	K. Ueda, J. Fujioka, B.J. Yang, J. Shiogai, A. Tsukazaki, S. Nakamura, S.
	Awaji, N. Nagaosa, and Y. Tokura, \emph{Magnetic field-induced
		insulator-semimetal transition in a pyrochlore
		{N}d$_{2}${I}r$_{2}${O}$_{7}$}, Phys. Rev. Lett. 115 (2015), p. 056402.
	
	\bibitem{Nakayama2016}
	M. Nakayama, T. Kondo, Z. Tian, J.J. Ishikawa, M. Halim, C. Bareille, W.
	Malaeb, K. Kuroda, T. Tomita, S. Ideta, K. Tanaka, M. Matsunami, S. Kimura,
	N. Inami, K. Ono, H. Kumigashira, L. Balents, S. Nakatsuji, and S. Shin,
	\emph{Slater to mott crossover in the metal to insulator transition of
		{Nd$_{2}$Ir$_2$O$_{7}$}}, Phys. Rev. Lett. 117 (2016), p. 056403.
	
	\bibitem{Nagamiya1982}
	T. Nagamiya, S. Tomiyoshi, and Y. Yamaguchi, \emph{Triangular spin
		configuration and weak ferromagnetism of {Mn$_3$Sn} and {Mn$_3$Ge}}, Solid
	State Commun. 42 (1982), pp. 385--388.
	
	\bibitem{Tomiyoshi1982}
	S. Tomiyoshi and Y. Yamaguchi, \emph{Polarized {N}eutron {D}iffraction {S}tudy
		of the {S}pin {S}tructure of {Mn$_3$Sn}}, J. Phys. Soc. Jpn. 51 (1982), pp.
	803--810.
	
	\bibitem{Tomiyoshi1983}
	S. Tomiyoshi, Y. Yamaguchi, and T. Nagamiya, \emph{Triangular spin
		configuration and weak ferromagnetism of {Mn$_3$Ge}}, J. Magn. Magn. Mater.
	31 - 34, Part 2 (1983), pp. 629--630.
	
	\bibitem{Kubler2014}
	J. K\"ubler and C. Felser, \emph{Non-collinear antiferromagnets and the
		anomalous {H}all effect}, Europhys. Lett. 108 (2014), p. 67001.
	
	\bibitem{hurd}
	C.M. Hurd, \emph{The {H}all {E}ffect in {M}etals and {A}lloys}, Plenum, New
	York, 1972.
	
	\bibitem{Burkov2011}
	A.A. Burkov and L. Balents, \emph{Weyl {S}emimetal in a {T}opological
		{I}nsulator {M}ultilayer}, Phys. Rev. Lett. 107 (2011), p. 127205.
	
	\bibitem{Yang2017njp}
	H. Yang, Y. Sun, Y. Zhang, W.J. Shi, S.S.P. Parkin, and B. Yan,
	\emph{Topological {Weyl} semimetals in the chiral antiferromagnetic materials
		{Mn}$_3${Ge} and {Mn}$_3${Sn}}, New Journal of Physics 19 (2017), p. 015008.
	
	\bibitem{Nielsen83plb}
	H.B. Nielsen and M. Ninomiya, \emph{The {A}dler-{B}ell-{J}ackiw anomaly and
		{W}eyl fermions in a crystal}, Physics Letters B 130 (1983), pp. 389--396.
	
	\bibitem{Son2013}
	D.T. Son and B.Z. Spivak, \emph{Chiral anomaly and classical negative
		magnetoresistance of {W}eyl metals}, Phys. Rev. B 88 (2013), p. 104412.
	
	\bibitem{Xiong2015}
	J. Xiong, S.K. Kushwaha, T. Liang, J.W. Krizan, M. Hirschberger, W. Wang, R.J.
	Cava, and N.P. Ong, \emph{Evidence for the chiral anomaly in the {D}irac
		semimetal {Na}$_3${Bi}}, Science 350 (2015), p. 413.
	
	\bibitem{Zhang2016}
	C.L. Zhang, S.Y. Xu, I. Belopolski, Z. Yuan, Z. Lin, B. Tong, G. Bian, N.
	Alidoust, C.C. Lee, S.M. Huang, T.R. Chang, G. Chang, C.H. Hsu, H.T. Jeng, M.
	Neupane, D.S. Sanchez, H. Zheng, J. Wang, H. Lin, C. Zhang, H.Z. Lu, S.Q.
	Shen, T. Neupert, M. Zahid~Hasan, and S. Jia, \emph{Signatures of the
		{A}dler-{B}ell-{J}ackiw chiral anomaly in a {W}eyl fermion semimetal}, Nature
	Commun. 7 (2016), p. 10735.
	
	\bibitem{Hirschberger16naturemat}
	M. Hirschberger, S. Kushwaha, Z. Wang, Q. Gibson, C.A. Belvin, B. Bernevig, R.
	Cava, and N. Ong, \emph{The chiral anomaly and thermopower of {W}eyl fermions
		in the half-{H}eusler {G}d{P}t{B}i}, Nature Materials 15 (2016), pp.
	1161--1165.
	
	\bibitem{Fukuyama}
	H. Fukuyama, \emph{Anomalous orbital magnetism and {H}all effect of massless
		fermions in two dimension}, J. Phys. Soc. Jpn. 76 (2007), p. 043711.
	
	\bibitem{Niu2010}
	D. Xiao, M.C. Chang, and Q. Niu, \emph{Berry phase effects on electronic
		properties}, Rev. Mod. Phys. 82 (2010), pp. 1959--2007.
	
	\bibitem{MacDonald2006}
	A.S. N\'u\~nez, R.A. Duine, P. Haney, and A.H. MacDonald, \emph{Theory of spin
		torques and giant magnetoresistance in antiferromagnetic metals}, Phys. Rev.
	B 73 (2006), p. 214426.
	
	\bibitem{Jungwirth2010}
	A.B. Shick, S. Khmelevskyi, O.N. Mryasov, J. Wunderlich, and T. Jungwirth,
	\emph{Spin-orbit coupling induced anisotropy effects in bimetallic
		antiferromagnets: {A} route towards antiferromagnetic spintronics}, Phys.
	Rev. B 81 (2010), p. 212409.
	
	\bibitem{MacDonald2011}
	A.H. MacDonald and M. Tsoi, \emph{Antiferromagnetic metal spintronics}, Phil.
	Trans. R. Soc. A 369 (2011), pp. 3098--3114.
	
	\bibitem{Park2011}
	B.G. Park, J. Wunderlich, X. Marti, V. Holy, Y. Kurosaki, M. Yamada, H.
	Yamamoto, A. Nishide, J. Hayakawa, H. Takahashi, A.B. Shick, and T.
	Jungwirth, \emph{A spin-valve-like magnetoresistance of an
		antiferromagnet-based tunnel junction}, Nature Mater. 10 (2011), pp.
	347--351.
	
	\bibitem{Marti2014}
	X. Marti, I. Fina, C. Frontera, J. Liu, P. Wadley, Q. He, R.J. Paull, J.D.
	Clarkson, J. Kudrnovsk{\'y}, I. Turek, J. Kune{\v{s}}, D. Yi, J.H. Chu, C.T.
	Nelson, L. You, E. Arenholz, S. Salahuddin, J. Fontcuberta, T. Jungwirth, and
	R. Ramesh, \emph{Room-temperature antiferromagnetic memory resistor}, Nature
	Mater. 13 (2014), pp. 367--374.
	
	\bibitem{Gomonay2014}
	E.V. Gomonay and V.M. Loktev, \emph{Spintronics of antiferromagnetic systems},
	Low Temp. Phys. 40 (2014), pp. 17--35.
	
	\bibitem{Jungwirth2016}
	T. Jungwirth, X. Marti, P. Wadley, and J. Wunderlich, \emph{Antiferromagnetic
		spintronics}, Nature Nanotech. 11 (2016), pp. 231--241.
	
	\bibitem{Fert2007}
	C. Chappert, A. Fert, and F.N. Van~Dau, \emph{The emergence of spin electronics
		in data storage}, Nature Mater. 6 (2007), pp. 813--823.
	
\end{thebibliography}

\end{document}